\documentclass[cits]{PoS}

\usepackage{amsmath}

\newcommand{\tev}{\,\operatorname{TeV}}
\newcommand{\gev}{\,\operatorname{GeV}}

\newcommand{\ms}{\mskip 1.5mu}

\newcommand{\jpsi}{J\mskip -2mu/\mskip -0.5mu\Psi}

\renewcommand{\vec}[1]{\textbf{#1}}

\title{\raggedleft{\small DESY 13-111}  \\[2em] \raggedright
  Multiparton interactions: Theory and experimental findings}

\ShortTitle{Multiparton interactions: Theory and experimental findings}

\author{\speaker{Markus DIEHL}\\
        Deutsches Elektronen-Synchroton DESY, 22603 Hamburg, Germany \\
        E-mail: \email{markus.diehl@desy.de}}

\abstract{I give an introduction to multiparton interactions in
proton-proton collisions, with a focus on the perturbative regime.
Recent experimental results are discussed, as well as progress and
open questions in theory.}

\FullConference{XXI International Workshop on Deep-Inelastic Scattering
  and Related Subjects-DIS2013,\\
  22-26 April 2013\\
  Marseilles,France}

\begin{document}

\section{Introduction}

\begin{figure}[h]
\begin{center}
\includegraphics[width=0.43\textwidth]{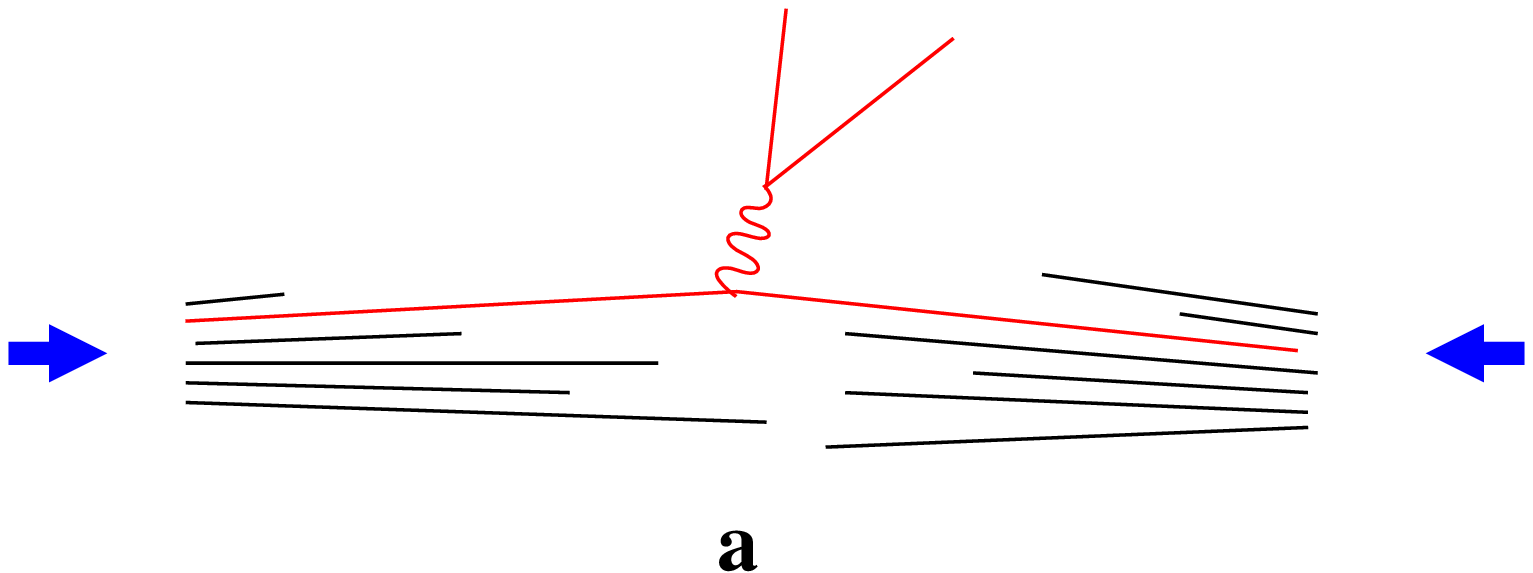}
\hspace{2em}
\includegraphics[width=0.43\textwidth]{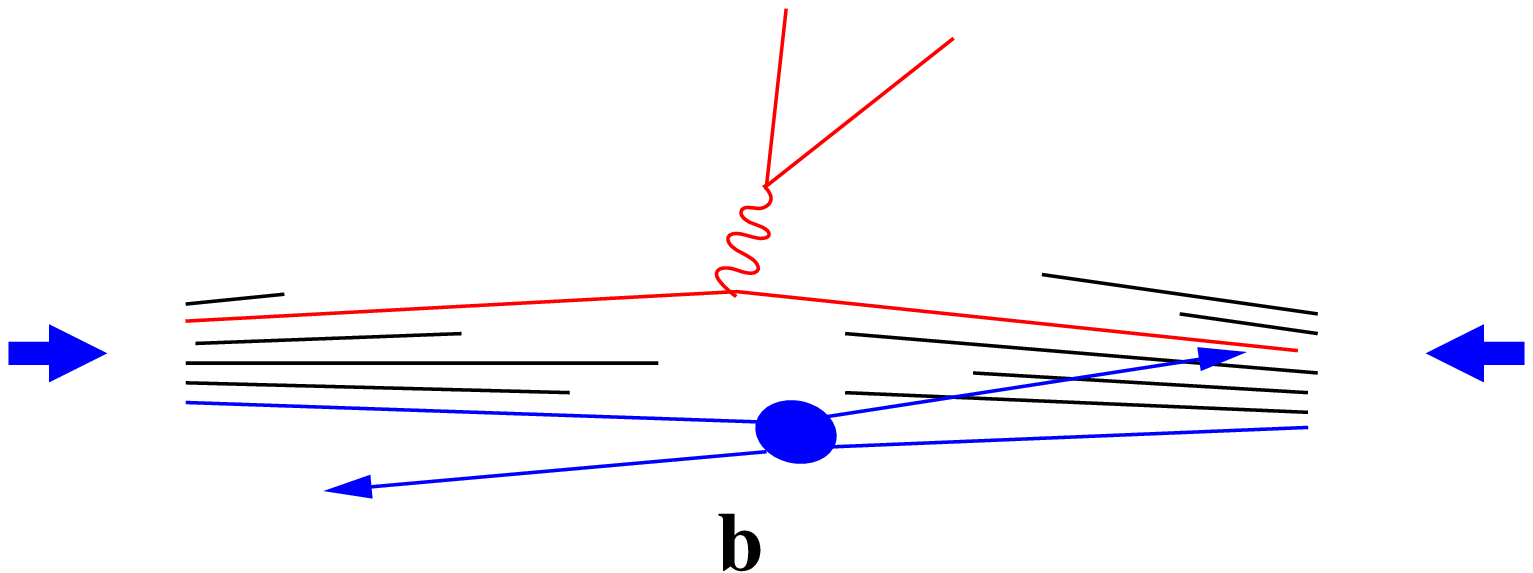}
\end{center}
\caption{\label{fig:dy-cartoons} Sketch of the process $pp\to Z + X \to
  \ell^+\ell^- + X$ without (a) and with (b) scattering amongst the
  ``spectator'' partons.}
\begin{center}
\includegraphics[width=0.39\textwidth]{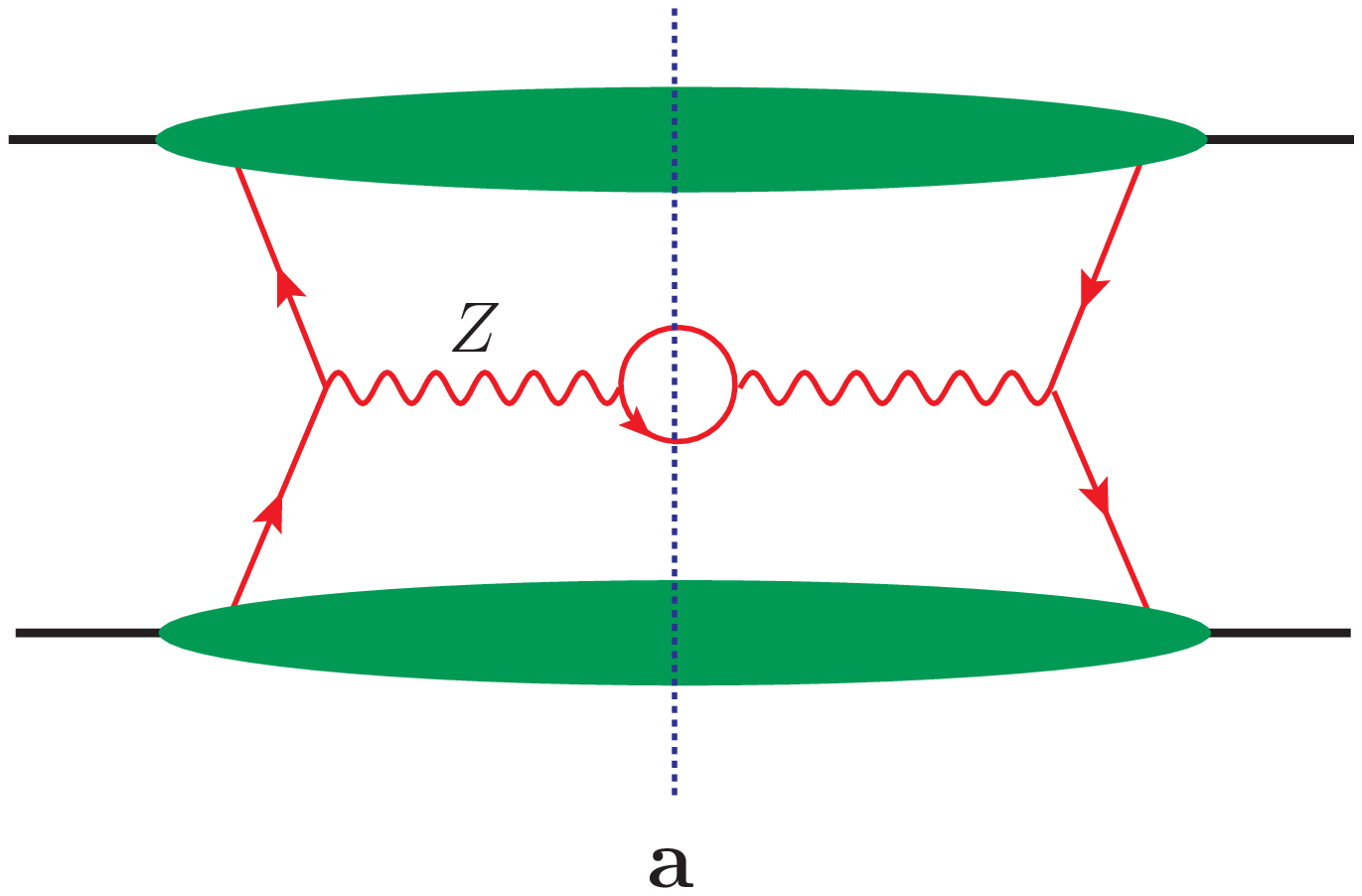}
\hspace{3em}
\includegraphics[width=0.39\textwidth]{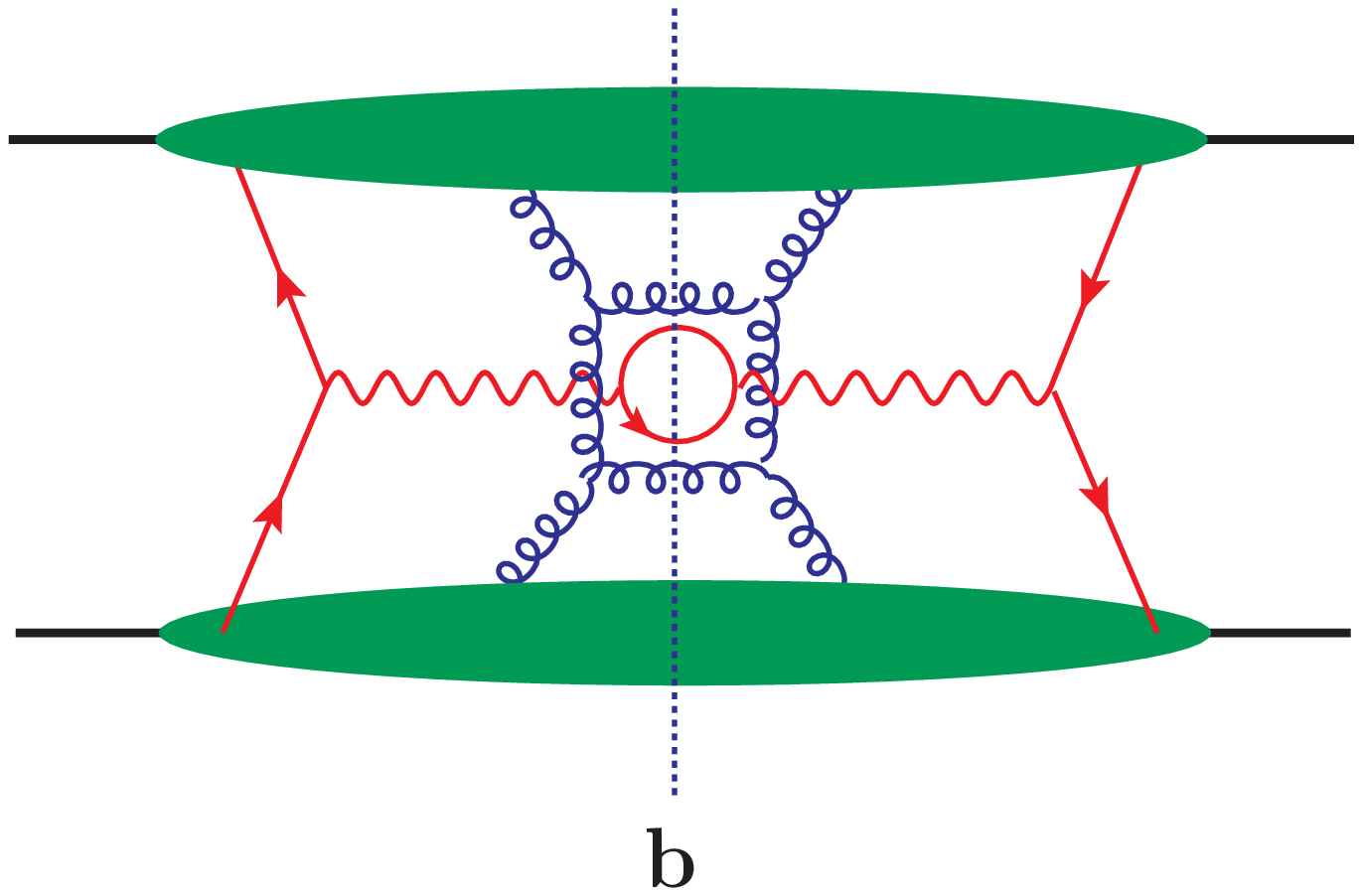}
\end{center}
\caption{\label{fig:dy-graphs} Cross section graphs corresponding to
  figure~\protect\ref{fig:dy-cartoons}.  The vertical line indicates the
  final state.}
\end{figure}

The theoretical basis for describing hard processes in proton-proton
collisions is provided by factorization formulae, which express an
observable cross section in terms of parton densities and cross sections
for hard-scattering subprocesses at parton level.  A textbook example is
the process $pp\to Z + X \to \ell^+\ell^- + X$, for which a sketch is
shown in figure \ref{fig:dy-cartoons}a and a graph in figure
\ref{fig:dy-graphs}a.  It is essential to realize that such factorization
formulae hold for \emph{inclusive} cross sections: they contain all
details of the particles produced in the hard-scattering subprocess (such
as the momenta of the leptons $\ell^+$ and $\ell^-$ in our example) but
give no information about the other particles (denoted by $X$ in the above
formula).  The physics picture suggested by figures \ref{fig:dy-cartoons}a
and \ref{fig:dy-graphs}a is deceptively simple, since it suggests that
only the two partons annihilating into a $Z$ boson interact.  This is not
the case: the other partons in the colliding protons interact with each
other as well, but the effects of their interactions cancel in the
inclusive cross section thanks to unitarity.  If, however, we ask for
details about the final state that concern the particles in $X$, these
interactions do matter.  An example is given in figures
\ref{fig:dy-cartoons}b and \ref{fig:dy-graphs}b.

Most frequently, these additional interactions produce particles that have
low transverse momentum and are part of the ``underlying event'' (UE),
which is defined as ``anything not produced in the hard-scattering
subprocess''.  (Clearly, this definition is not based on observable
features but depends on a specific theoretical description.)  At
high-energy colliders, especially the Tevatron and the LHC, the additional
interactions can, however, also be hard and produce particles with large
transverse momentum or large mass.  One then speaks of ``multiple hard
scattering''.  This talk is focused on the case with one additional hard
scattering, i.e.\ on ``double parton scattering'' (DPS).  The terms
``multiparton interactions'' (MPI) and ``multiple interactions'' are
sometimes restricted to the case where all interactions are hard and
sometimes meant to include the case where the additional interactions
contribute to the underlying event.

Double parton scattering can be an important background in searches for
new physics.  In the process $pp\to H Z + X \to b\bar{b}\, Z + X$, one
hard scatter can produce the $Z$ and the other a continuum $b\bar{b}$ pair
\cite{DelFabbro:1999tf}.  If the second scatter produces a Higgs boson,
one has a contribution to the signal.  The same holds for $pp\to H W + X
\to b\bar{b}\, W + X$, where the DPS contribution to the background has
been studied in detail for Tevatron kinematics in \cite{Bandurin:2010gn}.
A process with DPS contributions of prominent size is $W^+ W^+$ or $W^-
W^-$ production \cite{Kulesza:1999zh,Gaunt:2010pi}, which is a search
channel for supersymmetry \cite{Baer:2013yha}.

A general statement about the relative size of single and double parton
scattering for a given final state can be made on the basis of power
counting \cite{Diehl:2011tt,Diehl:2011yj,Blok:2011bu}.  Let $Q$ be the
typical hard scale in a process and $\vec{p}_{T1}$, $\vec{p}_{T2}$ the net
transverse momenta of the particles produced in the first and the second
interaction of the DPS process, respectively.  (In the above example,
$\vec{p}_{T1}$ is the transverse momentum of the $b\bar{b}$ pair and
$\vec{p}_{T2}$ the transverse momentum of the $Z$.)  One then finds
\begin{align}
\frac{d\sigma_{\text{single}}}{d^2\vec{p}_{T1}\,
   d^2\vec{p}_{T2}} \sim 
\frac{d\sigma_{\text{double}}}{d^2\vec{p}_{T1}\,
   d^2\vec{p}_{T2}} \sim \frac{1}{\Lambda^2 \ms Q^4} \,,
\end{align}
where $\Lambda$ represents a typical hadronic scale.  Double parton
scattering is hence \emph{not} power suppressed in the differential cross
section.  However, the phase space of DPS is limited to $p_{T1} \sim
\Lambda$ and $p_{T2} \sim \Lambda$, whereas in single hard scattering we
have the limitation $|\vec{p}_{T1} + \vec{p}_{T2}| \sim \Lambda$ but
$p_{T1}$ and $p_{T2}$ can each be of order $Q$.  Integrating over the
transverse momenta, we thus obtain
\begin{align}
\sigma_{\text{single}} \sim \frac{1}{Q^2} ~\gg~
  \sigma_{\text{double}} \sim \frac{\Lambda^2}{Q^4} \,.
\end{align}
Double parton scattering is now power suppressed, in accordance with the
usual factorization formulae that describe only single hard scattering.
However, even in this situation DPS can be important.  In particular, its
contribution grows with the overall collision energy since for decreasing
parton momentum fractions $x$ the two-parton density grows faster than the
one-parton density.


\section{The cross section for double hard scattering}

It is natural to assume that double hard scattering can be described by
factorization formulae akin to those for single hard scattering.  At tree
level one then has
\begin{align}
  \label{dps-fact}
 \frac{d\sigma_{\text{double}}}{dx_1\, d\bar{x}_1\;
  dx_2\, d\bar{x}_2}
&= \frac{1}{C}\, 
\hat{\sigma}_1(x_1 \bar{x}_1)\, \hat{\sigma}_2(x_2\ms \bar{x}_2)
\int\! d^2\vec{b}\,\,
F(x_1, x_2, \vec{b}) \, F(\bar{x}_1, \bar{x}_2, \vec{b}) \,,
\end{align}
where $C$ is a combinatorial factor and $\hat{\sigma}_i$ the parton-level
cross section for subprocess $i$ ($=1,2$).  For simplicity we have not
displayed labels that specify the different partons.  The momentum
fractions $x_i$ and $\bar{x}_i$ can be reconstructed from the final-state
kinematics, whereas the transverse distance $\vec{b}$ between the two
hard-scattering processes is unobservable.  $F(x_1, x_2, \vec{b})$ is a
double parton distribution (DPD) and describes the probability density for
finding two partons with respective momentum fractions $x_1$ and $x_2$ at
a transverse distance $\vec{b}$ inside the proton.  It is sometimes useful
to Fourier transform $F(x_1, x_2, \vec{b})$ w.r.t.\ $\vec{b}$.  The
resulting distributions have been termed ``generalized two-parton
distributions'' in \cite{Blok:2011bu,Blok:2010ge}, which should be
distinguished from the generalized (single) parton distributions that
appear in exclusive processes such as $ep\to ep\gamma$
\cite{Sabatie:DIS2013}.

The formula \eqref{dps-fact} can be extended to include radiative
corrections in $\hat{\sigma}_i$ (with convolution integrals instead of
products between $\hat{\sigma}_i$ and $F$ on the r.h.s.) and to be
differential in the momenta of the final state particles produced by hard
scattering.  Notice that \eqref{dps-fact} still describes an
\emph{inclusive} cross section, i.e.\ it includes all further interactions
between ``spectator'' partons, which are predominantly soft but may
occasionally be hard.  We will comment on the theoretical status of
\eqref{dps-fact} in section~\ref{sec:fact}.

To use \eqref{dps-fact} for estimating DPS cross sections, one needs an
ansatz for the double parton distributions, which are essentially unknown.
\emph{If} one assumes that DPDs factorize as $F(x_1,x_2,\vec{b}) =
f(x_1)\, f(x_2)\, G(\vec{b})$, where $f(x_i)$ are the usual parton
densities, and \emph{if} one assumes that the transverse distance
distribution $G(\vec{b})$is the same for all parton types, then the cross
section formula \eqref{dps-fact} (as well as its generalization to higher
orders) turns into
\begin{align}
  \label{pocket-form}
 \frac{d\sigma_{\text{double}}}{dx_1\, d\bar{x}_1\;
  dx_2\, d\bar{x}_2}
&= \frac{1}{C}\,
   \frac{d\sigma_1}{dx_1\, d\bar{x}_1} \;
   \frac{d\sigma_2}{dx_2\, d\bar{x}_2}\;
       \frac{1}{\sigma_{\text{eff}}}
\end{align}
with cross sections $\sigma_{i}$ for single hard scattering and a
universal factor $1 /\sigma_{\text{eff}} = \textstyle{\int} d^2\vec{b}\;
\bigl[ G(\vec{b}) \bigr]{}_{}^2$.  With this ``pocket formula'' and a
value for $\sigma_{\text{eff}}$ in hand, one can conveniently estimate the
rate for double parton scattering in any given process.  One should,
however, bear in mind that \eqref{pocket-form} relies on strong
simplifications and must be expected to have a limited accuracy.
Alternatively, one may use \eqref{pocket-form} to define
$\sigma_{\text{eff}}$ and extract it from experiment.  A dependence of
$\sigma_{\text{eff}}$ on the process or on kinematic variables then
indicates that the assumptions on $F(x_1,x_2,\vec{b})$ spelled out before
\eqref{pocket-form} are too simple.


\section{Recent experimental results}

It has long been known that multiple interactions are indispensable for
describing the underlying event in $pp$, $p\bar{p}$ and $\gamma p$
collisions (see \cite{Sjostrand:2004pf} for an earlier review).  It is
hence not surprising that experimental studies of the underlying event can
be used to tune those parameters in Monte Carlo event generators that
describe the modeling of multiple interactions.  We will not expand on
this subject here but refer to the presentations
\cite{Kar:DIS2013,Mazumdar:DIS2013,Kepka:DIS2013} at this workshop.

\begin{figure}
\begin{center}
\includegraphics[height=0.38\textwidth]{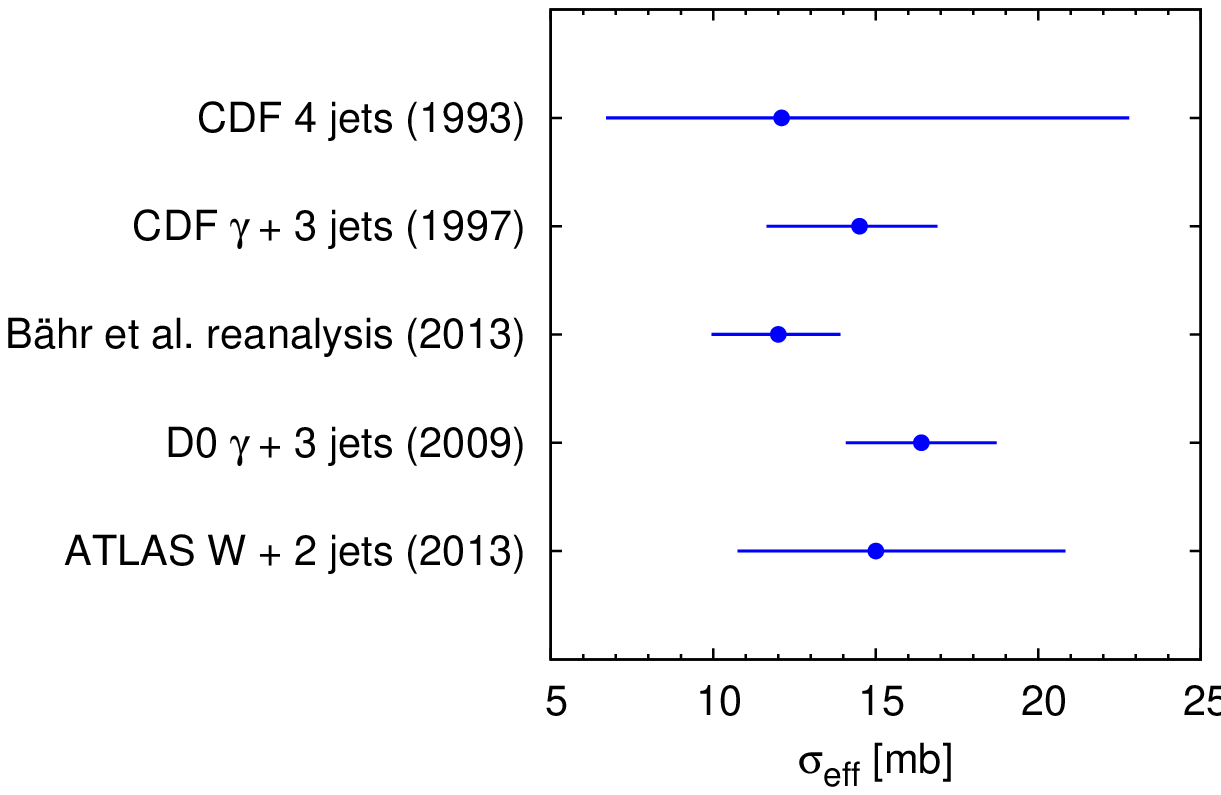}
\hspace{0.5em}
\includegraphics[height=0.38\textwidth]{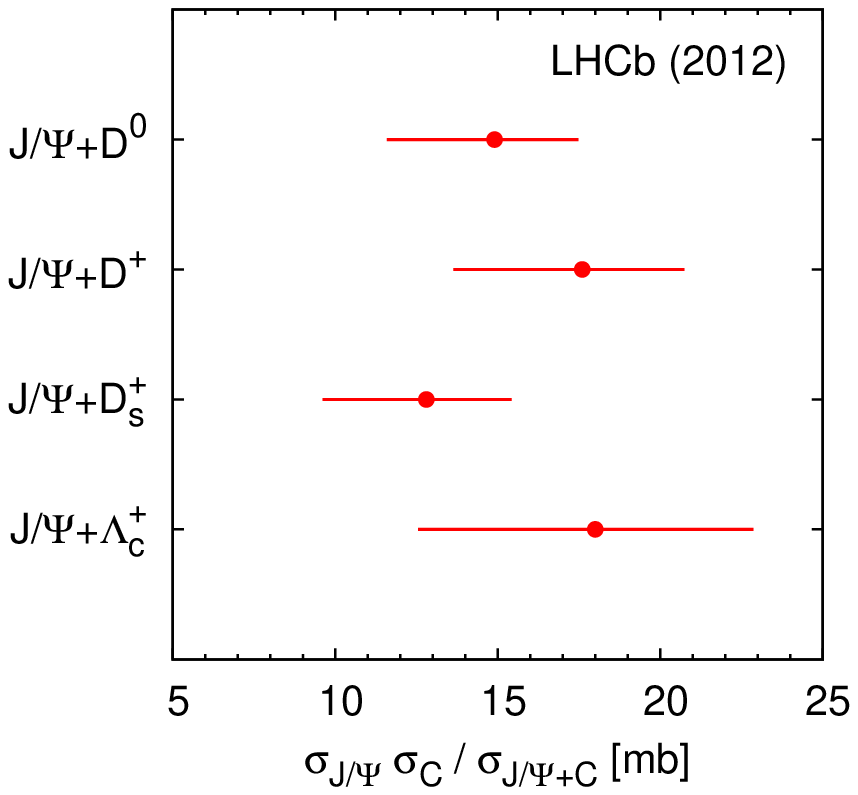}
\end{center}
\caption{\label{fig:sigma-eff} Left: experimental determinations of
  $\sigma_{\text{eff}}$ by CDF \protect\cite{Abe:1993rv,Abe:1997xk}, D0
  \protect\cite{Abazov:2009gc} and ATLAS \protect\cite{Aad:2013bjm}, as
  well as a reanalysis of the CDF data \protect\cite{Abe:1997xk} by B\"ahr
  et al.~\protect\cite{Bahr:2013gkj}.  Right: cross section ratios
  measured by LHCb \protect\cite{Aaij:2012dz} for inclusive production of
  a $\jpsi$, a charmed hadron $C$, and a pair $\jpsi + C$.  In both
  panels, statistical and systematic uncertainties have been added in
  quadrature.}
\end{figure}

In the left panel of figure~\ref{fig:sigma-eff} we collect determinations
of $\sigma_{\text{eff}}$ from several experimental studies at the Tevatron
and the LHC.  Within the uncertainties, there is no clear indication for a
dependence on the process or the collision energy.  The D0 analysis
\cite{Abazov:2009gc} has also performed a differential extraction in three
bins for the $p_T$ of the second hardest jet; the results are consistent
with no dependence but also with a slight decrease of
$\sigma_{\text{eff}}$ with $p_T$.

It must be emphasized that these determinations of $\sigma_{\text{eff}}$
are very difficult, since all considered processes have significant
contributions from single hard scattering, which must be understood with
high precision in order to determine the contribution from DPS.  At this
workshop, both ATLAS \cite{Aad:2013bjm,Myska:DIS2013} and CMS
\cite{CMS:awa,Bartalini:DIS2013} have presented studies of DPS in the
production of a $W$ associated with exactly two jets having $p_T > 20
\gev$.  Among the variables used to exhibit DPS is the relative transverse
momentum imbalance
\begin{align}
\Delta_{\text{jets}}^{n}(\text{ATLAS})
= \Delta^{\text{rel}}p_T(\text{CMS})
&= \frac{|\vec{p}_{T\,\text{jet}1} + \vec{p}_{T\,\text{jet}2}|}{%
  |\vec{p}_{T\,\text{jet}1}| + |\vec{p}_{T\,\text{jet}2}|}
\end{align}
between the two jets, whose distribution in single hard scattering covers
a broad range between 0~and 1 but is peaked at small values in DPS.  The
ATLAS analysis \cite{Aad:2013bjm} extracted a fraction $f_{\text{DP}} =
0.08 \pm 0.01 \pm 0.02$ of double parton scattering in their event sample,
which is very significant for a process that is formally a power
correction to single hard scattering.

There are several investigations of multiparton interactions in the charm
sector.  Among those is an ALICE study of particle multiplicities in
$\jpsi$ production \cite{Abelev:2012rz}.  An earlier LHCb measurement
\cite{Aaij:2011yc} has raised significant interest since it suggests that
double parton parton scattering may play a prominent role in the
production of $\jpsi$ pairs \cite{Kom:2011bd,Novoselov:2011ff,%
  Baranov:2011ch,Baranov:2012re,Zotov:DIS2013}.  Even more spectacularly,
the cross sections measured by LHCb for the production of a $\jpsi$
associated with a charmed hadron $C$ are more than one order of magnitude
larger than predictions based on single hard scattering
\cite{Aaij:2012dz}.  If this process is dominated by double parton
scattering, the pocket formula \eqref{pocket-form} gives
$\sigma_{\text{eff}} = \sigma_{\jpsi}\, \sigma_{C\phantom{/}\!\!}
/\sigma_{\jpsi+C}$.  The values for this cross section ratio determined by
LHCb are shown in the right panel of figure~\ref{fig:sigma-eff}.  Their
size suggests that DPS may indeed dominate the production channels in
question.  The situation is, however, not so simple if one looks at
differential distributions.  The $p_T$ slopes of the charmed hadron in
$\jpsi+C$ production are comparable to those in prompt $C$ production, but
the $p_T$ spectrum of the $\jpsi$ in the $\jpsi+D^0$ and $\jpsi+D^+$
channels is significantly less steep than in inclusive $\jpsi$ production
(in the $\jpsi+D_s$ and $\jpsi+\Lambda_c$ channels, larger errors prevent
a similarly strong statement).  This does not support the simplest picture
of DPS where there is no correlation between the two partons in each
proton, so that the two hard scatters are completely independent of each
other.  It will be interesting to follow up on these issues, also for
double open charm production, which has been measured by LHCb as well
\cite{Aaij:2012dz}.  Theoretical analyses of these channels can be found
in \cite{Berezhnoy:2012xq,Luszczak:2011zp,Maciula:DIS2013}.


\section{A closer look at theory}

The dynamics of double parton scattering is far more involved than then
pocket formula \eqref{pocket-form} suggests, and we will now discuss a few
theoretical aspects that have been investigated in the recent literature.

\subsection{Parton correlations}

Beyond a certain precision, one cannot expect that the distribution
$F(x_1,x_2,\vec{b})$ of two partons in a proton is given by a product
$f(x_1)\, f(x_2)$ of usual PDFs times a universal factor $G(\vec{b})$.
Two-parton correlations involving $x_1, x_2$ and $\vec{b}$ lead to a more
complicated form of $F(x_1,x_2,\vec{b})$ and have been considered in
several studies.  Moreover, there can be correlations involving the
quantum numbers of the two partons, namely their polarization, color and
flavor.  This requires the addition of further distributions in the cross
section formula \eqref{dps-fact}.  More detail is given in
\cite{Diehl:DIS2013}.


\subsection{Parton splitting}

\begin{figure}[h]
\begin{center}
\vspace{0.3em}
\includegraphics[width=0.40\textwidth]{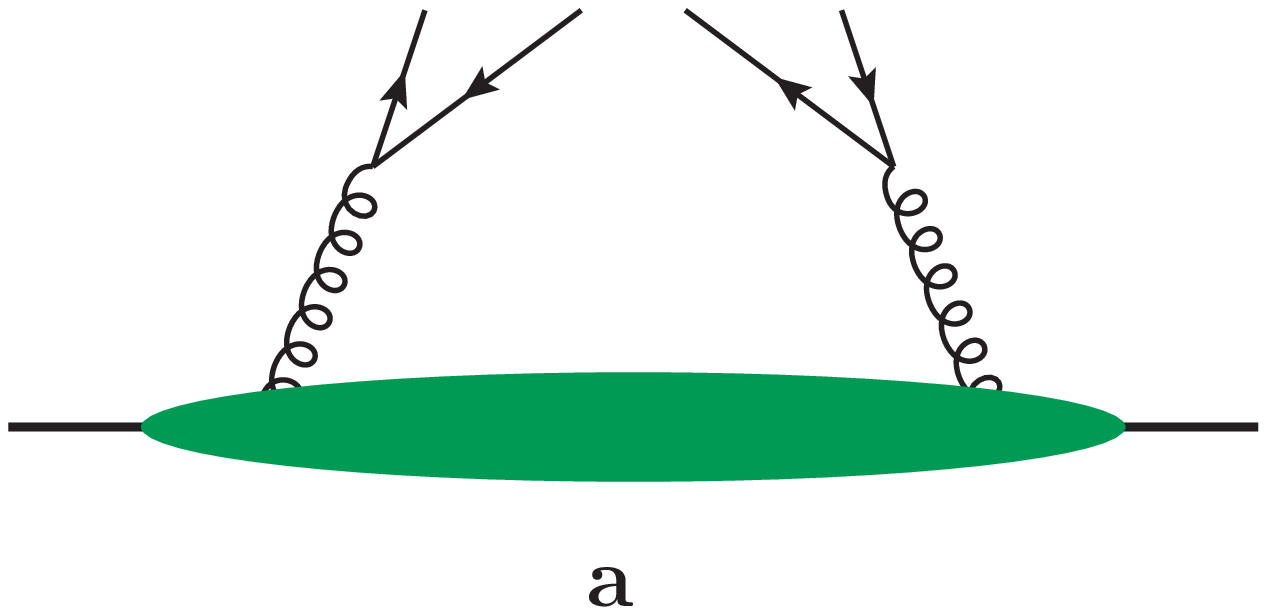}
\hspace{3em}
\includegraphics[width=0.40\textwidth]{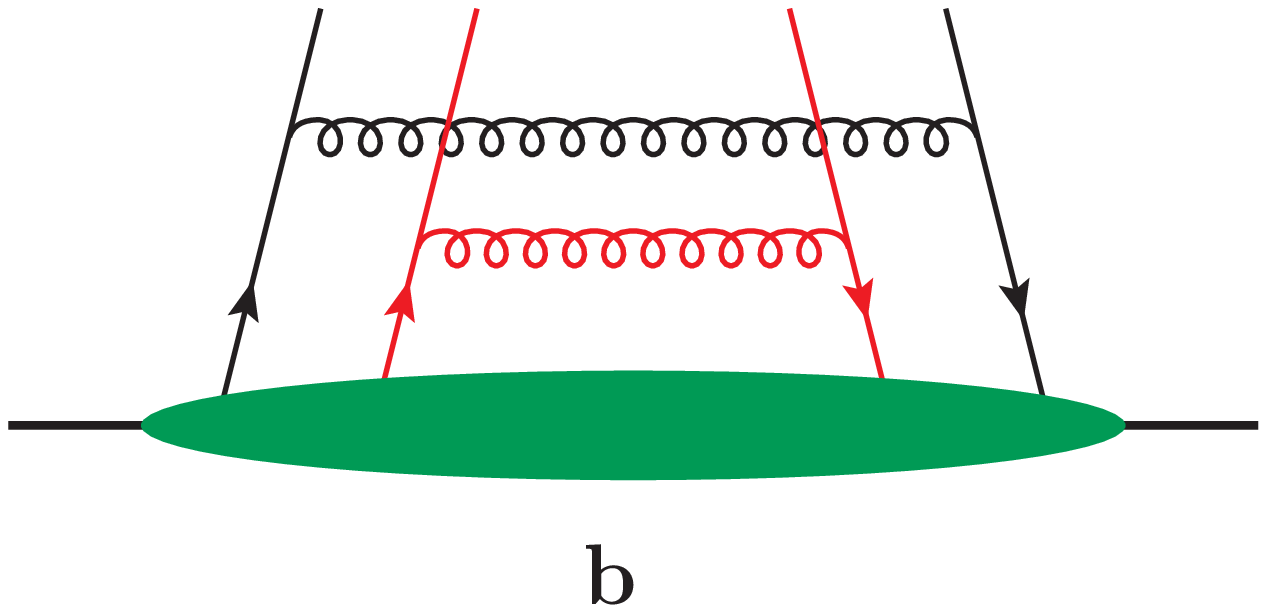}
\end{center}
\caption{\label{fig:evol} (a) A graph for the splitting of a single parton
  into two in a double parton distribution.  (b) A ladder graph for the
  separate DGLAP evolution of partons 1 and 2 in a double parton
  distribution.}
\end{figure}

While the typical transverse distance $\vec{b}$ between two partons in a
proton is of hadronic size, the cross section formula \eqref{dps-fact}
involves all values of $\vec{b}$ down to zero.  For sufficiently small
$\vec{b}$, one can compute DPDs in terms of PDFs and the perturbative
splitting of one parton into two \cite{Diehl:2011yj}.  At leading order in
$\alpha_s$, this gives
\begin{align}
  \label{dpd-splitting}
F(x_1,x_2, \vec{b}) &=  \frac{1}{\pi\ms \vec{b}^2}\;
  P\biggl(\frac{x_1}{x_1+x_2}\biggr)\; \frac{f(x_1+x_2)}{x_1+x_2} \,,
\end{align}
where $P$ is the familiar DGLAP splitting function for the relevant
process (again we omit labels for the parton species for simplicity). This
mechanism, shown in figure~\ref{fig:evol}a, generates substantial
correlations between $x_1$ and $x_2$ in the DPD, as well as between the
polarization and the color of the two partons.  It also plays a role in
the scale dependence of the DPDs.  One contribution to their evolution
equation describes the separate parton cascades radiated from parton 1 and
from parton 2, as shown in figure~\ref{fig:evol}b.  Depending on how
exactly DPDs are defined, their evolution equation contains or does not
contain an additional term originating from the splitting
\eqref{dpd-splitting}.  The evolution equation including this term has
been studied intensively
\cite{Kirschner:1979im,Shelest:1982dg,Snigirev:2003cq,%
  Gaunt:2009re,Ceccopieri:2010kg,Lewandowska:DIS2013}.  It remains,
however, to be established which version of evolution is actually relevant
in the factorization formula for DPS.

\begin{figure}
\begin{center}
\includegraphics[width=0.85\textwidth]{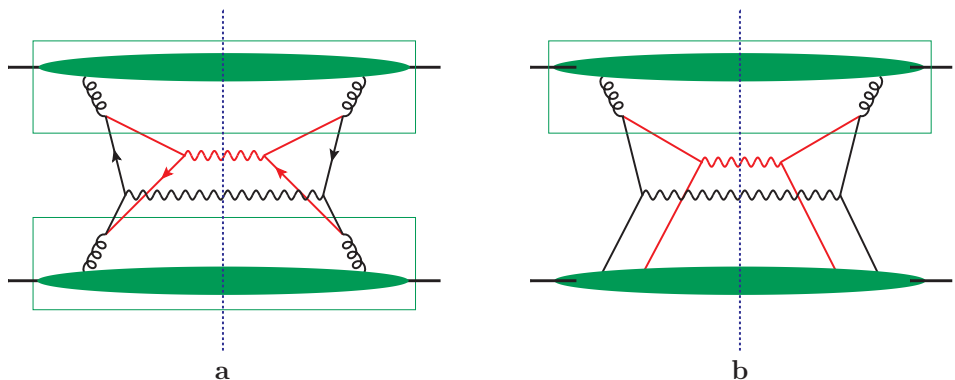}
\end{center}
\caption{\label{fig:split-X} Graphs for the production of two electroweak
  gauge bosons ($\gamma^*, Z, W$) involving the splitting of one parton
  into two in (a) both protons or (b) in one of the protons only.  The
  boxes indicate the small $\vec{b}$ limit of a DPD as shown in
  figure~\protect\ref{fig:evol}a.}
\end{figure}

The $1/\vec{b}^2$ short-distance singularity in \eqref{dpd-splitting}
gives an integral diverging like $\int d\vec{b}^2 /\vec{b}^4$ when
inserted in the cross section formula \eqref{dps-fact}.  This is obviously
inappropriate and indicates a deeper problem.  The graph in
figure~\ref{fig:split-X}a can be read as a contribution to DPS, with each
DPD being replaced by its leading contribution at small $\vec{b}$.  The
same graph is, however, a higher-order graph for the production of two
electroweak gauge bosons by gluon fusion.  To even define what is meant by
double parton scattering (and to derive a corresponding cross section
formula) requires a careful discussion of this double counting problem,
which was pointed out for multijet production already in
\cite{Cacciari:2009dp}.  Currently there is no consensus regarding its
solution \cite{Diehl:2011tt,Diehl:2011yj,Blok:2011bu,Gaunt:2011xd,%
  Ryskin:2011kk,Ryskin:2012qx}.

A similar discussion applies to the graph in figure~\ref{fig:split-X}b,
which can be read either as a hybrid of double and single parton
scattering, or as a contribution to DPS with the small $\vec{b}$ limit of
the DPD taken for the upper proton.  Detailed investigations of this
mechanism can be found in \cite{Blok:2010ge,Blok:2012mw,Blok:2013bpa} and
in \cite{Gaunt:2012dd}.


\subsection{Factorization}
\label{sec:fact}

\begin{figure}
\begin{center}
\includegraphics[width=0.43\textwidth]{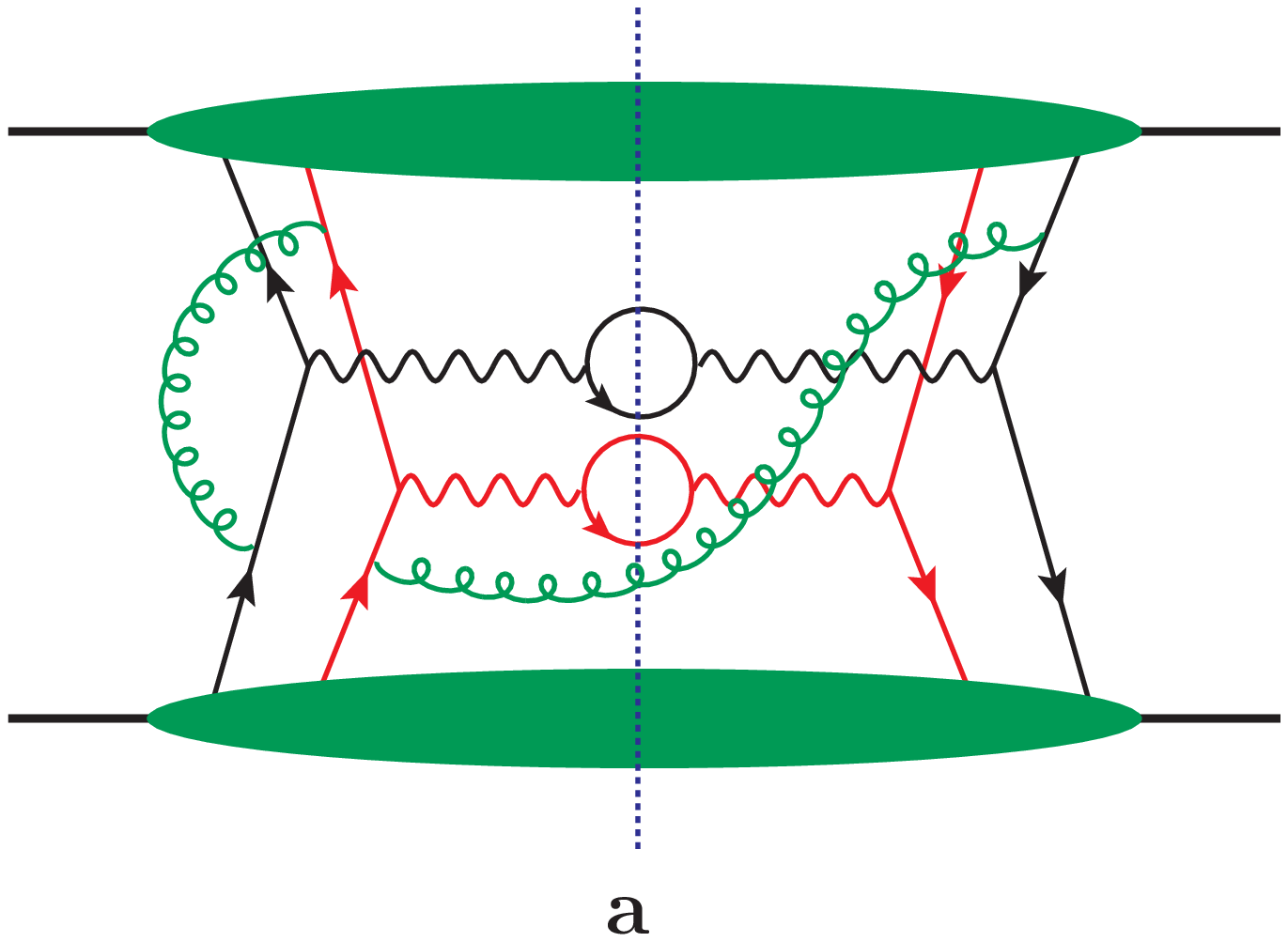}
\hspace{3em}
\includegraphics[width=0.43\textwidth]{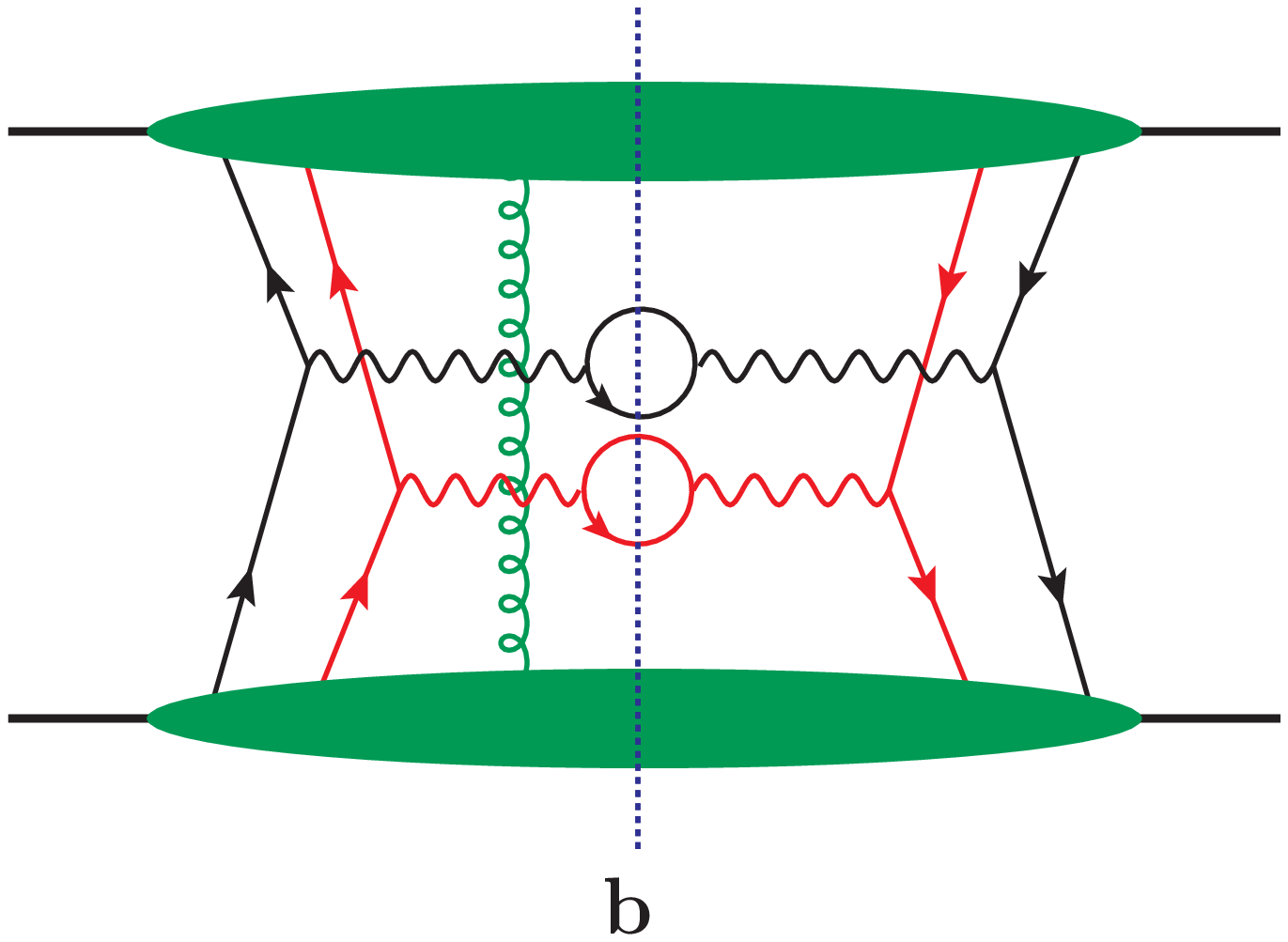}
\end{center}
\caption{\label{fig:factorization} Graphs for the production of two
  electroweak gauge bosons by double parton scattering with additional
  gluon exchange.  It is nontrivial to show that the effect of such graphs
  is properly taken into account in a factorization formula.}
\end{figure}

The factorization formula \eqref{dps-fact}, which forms the theory basis
for most estimates of double parton scattering, does not have the same
theoretical status as factorization formulae for single hard scattering in
hadron-hadron collisions.  At lowest perturbative order, i.e.\ for graphs
like those in figure~\ref{fig:dy-graphs}b, the formula can be derived
using standard techniques for approximating Feynman graphs
\cite{Paver:1982yp,Mekhfi:1983az,Diehl:2011yj}.  In particular, the
transverse distance $\vec{b}$ between the scattering partons naturally
appears in the quantum-level calculation, without any semiclassical
approximation.  To establish factorization requires many further elements,
regarding in particular the exchange of soft gluons between the different
parts of the graph.  Several ingredients to a possible factorization proof
have been given in \cite{Diehl:2011yj} and \cite{Manohar:2012jr}.  In
particular, one can show that soft-gluon exchange (see
figure~\ref{fig:factorization}) can be arranged into Sudakov factors if
the gluon momentum is in the kinematic region where the so-called eikonal
approximation can be made.

Among the questions that remain unsolved is whether soft gluon exchange
breaks factorization when the gluon momentum is in the so-called Glauber
region.  This is one of the most difficult parts in any factorization
proof for hadron-hadron collisions, and even for single hard scattering a
detailed argument only exists for a small number of processes.


\section{Conclusions}

This is a good time for gaining a deeper understanding of multiparton
interactions, and there is significant activity in the field
\cite{Bartalini:2011jp,Platzer:2012gla,Abramowicz:2013iva}.
The theoretical analysis has made important progress in the last few
years, but difficult questions still await an answer.  On the experimental
side, the jump in energy from the Tevatron to LHC has opened many new
possibilities for a detailed study of processes where double parton
scattering is expected to play a role.  First results from the LHC have
demonstrated that corresponding analyses are feasible even if they are not
easy.  Input from experiment will be essential to guide the development of
phenomenology.

Many recent results on multiparton interactions were not discussed in this
talk for reasons of time.  Among them are estimates of DPS contributions
to a variety of processes \cite{Maina:2009vx,Maina:2009sj,Maina:2010vh,%
  Berger:2009cm,Berger:2011ep,Kom:2011nu}, multiparton interactions in
proton-nucleus and nucleus-nucleus collisions
\cite{Calucci:2010wg,Strikman:2010bg,Blok:2012jr,%
  d'Enterria:2012qx,d'Enterria:2013ck,Treleani:2012zi}, and the small-$x$
approach \cite{Flensburg:2011kj,Bartels:2011qi,Cazaroto:2013fua}.  The
latter provides in particular a connection between multiparton
interactions and diffraction \cite{Bartels:2005wa}, as well as an
explanation of the ``ridge-effect'' in $pp$ and $pA$ collisions, which has
been observed by CMS, ATLAS and ALICE
\cite{Khachatryan:2010gv,Li:2012hc,CMS:2012qk,Aad:2012gla,Abelev:2012cya}.


\end{document}